\documentclass[sigconf]{acmart}
\usepackage{float}
\usepackage{xcolor,colortbl} % column color in table
\usepackage{amsmath,amsfonts} %
\usepackage{graphicx}%
\usepackage{textcomp}%
\usepackage{xcolor}%
\usepackage{enumitem}

\usepackage{array}%
\usepackage{pifont}%

\usepackage{amsthm}%
\usepackage{bm}

\usepackage[linesnumbered,ruled,vlined]{algorithm2e}%
\usepackage{setspace}%

\usepackage{multirow}%
\usepackage{siunitx}
\usepackage{footnote}%
\usepackage[utf8]{inputenc}
\usepackage[english]{babel}

\usepackage{blindtext}

\usepackage{dblfloatfix}%

\usepackage{xcolor,colortbl} % column color in table

\usepackage{xurl}
\usepackage{ctable} % for \specialrule command 
\graphicspath{ {pdf/} }

\usepackage{pifont}

\usepackage{filecontents}
\usepackage{float}

\newlength{\bibitemsep}
\newlength{\bibparskip}
\let\oldthebibliography\thebibliography
\renewcommand\thebibliography[1]{%
	\oldthebibliography{#1}%
	\setlength{\parskip}{\bibitemsep}%
	\setlength{\itemsep}{\bibparskip}%
}

\AtBeginDocument{%
  \providecommand\BibTeX{{%
    \normalfont B\kern-0.5em{\scshape i\kern-0.25em b}\kern-0.8em\TeX}}}

\setcopyright{acmcopyright}
\copyrightyear{2022}
\acmYear{2022}
\acmDOI{XXXXXXX.XXXXXXX}

\acmConference[41st ICCAD]{41st International Conference on Computer-Aided Design}{30 October--3 November, 2022}{San Diego, CA}
\setcopyright{none}
\renewcommand\footnotetextcopyrightpermission[1]{}

\begin{document}

%%
%% The "title" command has an optional parameter,
%% allowing the author to define a "short title" to be used in page headers.
\title{Sensor Security: Current Progress, Research Challenges, and Future Roadmap (Invited Paper)}

%%
%% The "author" command and its associated commands are used to define
%% the authors and their affiliations.
%% Of note is the shared affiliation of the first two authors, and the
%% "authornote" and "authornotemark" commands
%% used to denote shared contribution to the research.
%(Invited Paper)
\author{Anomadarshi Barua and Mohammad Abdullah Al Faruque}
%\authornote{Both authors contributed equally to this research.}
\affiliation{%
  \institution{Department of Electrical Engineering and Computer Science}
  \city{University of California, Irvine}
  \state{California}
  \country{USA.}\\
  Email: \textit{\{anomadab, alfaruqu\}@uci.edu}
}

%\textit{\{modema, nafiulr, bdemirel, alfaruqu\}}@uci.edu}\vspace{-8truemm}}

%%
%% By default, the full list of authors will be used in the page
%% headers. Often, this list is too long, and will overlap
%% other information printed in the page headers. This command allows
%% the author to define a more concise list
%% of authors' names for this purpose.
\renewcommand{\shortauthors}{Barua et al.}

%%
%% The abstract is a short summary of the work to be presented in the
%% article.
\begin{abstract}

Sensors are one of the most pervasive and integral components of today's safety-critical systems. Sensors serve as a bridge between physical quantities and connected systems.  The connected systems with sensors blindly believe the sensor as there is no way to authenticate the signal coming from a sensor. This could be an entry point for an attacker. An attacker can inject  a fake input signal along with the legitimate signal by using a suitable spoofing technique. As the sensor's transducer is not smart enough to differentiate between a fake and legitimate signal, the injected fake signal eventually can collapse the connected system. This type of attack is known as the transduction attack. Over the last decade, several works have been published to provide a defense against the transduction attack. However, the defenses are proposed on an ad-hoc basis; hence, they are not well-structured. Our work begins to fill this gap by providing a checklist that a  defense technique should always follow to be considered as an ideal defense against the transduction attack. We name this checklist as the \textit{Golden reference} of sensor defense. We provide insights on how this Golden reference can be achieved and argue that sensors should be redesigned from the transducer level to the sensor electronics level. We point out that only  hardware or  software modification is not enough; instead, a hardware/software (HW/SW)  co-design approach is required to ride on this future roadmap to the robust and resilient sensor.

\end{abstract}

%%
%% The code below is generated by the tool at http://dl.acm.org/ccs.cfm.
%% Please copy and paste the code instead of the example below.
%%


%%
%% Keywords. The author(s) should pick words that accurately describe
%% the work being presented. Separate the keywords with commas.

%% A "teaser" image appears between the author and affiliation
%% information and the body of the document, and typically spans the
%% page.

%%
%% This command processes the author and affiliation and title
%% information and builds the first part of the formatted document.
\maketitle

\section{Introduction}
\label{sec:Introduction}

Sensors work as the eyes and ears of any embedded system. Therefore, any decision taken by a system depends upon the data coming from a sensor. Though sensors are technically developed nowadays compared to their earlier generation, from a security point of view, almost all sensors are still unsafe and prone to intelligent attacks by an attacker. The reason behind this is that the signals that a sensor measures in the surrounding environment are analog signals, which are not encrypted in the environment. Therefore, sensors cannot differentiate between a legitimate analog input and a fake analog input provided by an attacker. As a result, the transducer of a sensor converts any fake analog signal given as the input into a usable signal (e.g., an electrical signal), which is then propagated to downstream of the signal path and eventually arrives at the system controller. As the system controller does not have any way to authenticate the signal coming from a sensor, it has to blindly believe the sensor and then make decision based upon the sensor data. As there is no hardware/software firewall present between the sensor and the system interface, this could be an entry point for an attacker. An attacker can \textit{noninvasively} inject fake signals into sensors by using a suitable spoofing technique \cite{barua2020hallspoofing, yan2016can, zhang2017dolphinattack, wang2017sonic} and may compromise the system availability and integrity, causing  a system failure and denial-of-service (DoS) attacks on systems.

As the attack at hand fools the sensor's transducer, this type of attack is named as the \textit{transduction attack} \cite{fu2018risks}. There has been ongoing research on the transduction attack for the last two decades. However, this paper shows that defenses against this transduction attack are still not well-structured, and  little to no work is done in the sensor defense domain. We point out that almost all defenses against the transduction attack are proposed on an ad-hoc basis. Therefore, there is no systematic approach present in the literature for defense. Our work begins to fill this gap by providing a checklist of reference points that an ideal defense should always follow to be considered as a successful defense against the transduction attack. We name this checklist as the \textit{Golden reference} for sensor defense. 

This Golden reference is the roadmap to a secured sensor that every designer should follow in the future before finally adopting a technique as a sensor defense. We also provide insights on how to achieve the Golden reference. We point out that the Golden reference for sensor defense can not be achieved alone by only  hardware or only  software modification; instead, a hardware/software (HW/SW) co-design approach is required in the sensor domain. The sensors should be redesigned from the
transducer level to the sensor electronics level with the integration of smart hardware and software. We believe that the defenses highlighted in this paper will be relevant soon when sensors will pervade our lives.

\section{Background}
\label{sec:Background}

%A brief background is provided below to facilitate the understanding of the sensors and their associated vulnerability.

\subsection{Sensor physics}
\label{subsec:Sensor physics}

A sensor is an instrument that senses physical input signals from the environment and converts the measured physical quantities into understandable data that can be interpreted by either a human or a machine. For example, a simple glass thermometer presents visual data, which can be interpreted by a human, and an electronic pressure sensor can provide electronic data, which can be interpreted by machines. There are a multitude of types of sensors and the physics behind the conversion of physical input signals to  understandable data varies from one sensor to another. However, all sensors have a common physics that is systematized below.

\begin{figure}[ht!]
\centering
\includegraphics[width=0.475\textwidth,height=0.08\textheight]{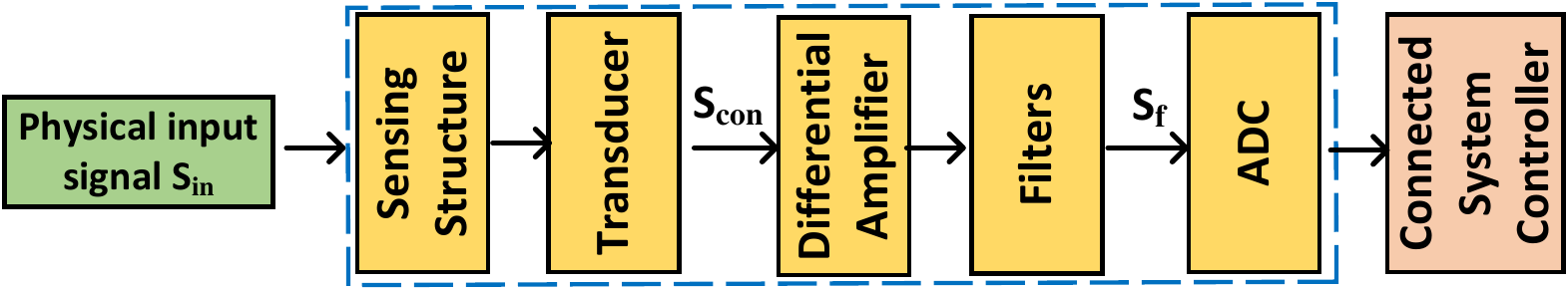}
\vspace{-2.20em}
\caption{A basic overview of the sensor physics.}
\label{fig:sensor_structure}
\vspace{-01.310em}
\end{figure}

\textbf{Sensing structure and transducer:} The most common sensor physics is that all the sensors use a sensing structure, which typically senses the physical input signals and uses a \textit{transducer} to convert the sensed physical input signals into understandable data (see Fig. \ref{fig:sensor_structure}). For example, a pressure sensor has a diaphragm as the sensing structure and uses a capacitor plate as the transducer to convert the displacement of the diaphragm into an electrical signal. The converted understandable data is ideally proportional to physical input signals. Mathematically, if the sensed physical input signal is $S_{in}$ and the corresponding converted data from the transducer is $S_{con}$, we can write $S_{con}$ as follows:

\vspace{-0.700em}
\begin{equation}
S_{con} = k \times f\{S_{in}\}
\label{eqn:sensor_equation}
\end{equation}
\vspace{-1.0em}

where $k$ is a proportionality constant and $f\{S_{in}\}$ is a function of the $S_{in}$. It is important to note that $f\{S_{in}\}$ can be a function of any type. For example, for a Hall sensor, $f\{S_{in}\}$ is a linear function of input magnetic fields and for a p-n junction temperature sensor, $f\{S_{in}\}$ is an exponential function of input temperature $T$ (i.e., $e^T$). 

\vspace{-1.00em}

\subsection{Sensor electronics}
\label{subsec:Sensor electronics}

Though sensing structure and transducer are common in all sensors, most sensors have other additional electronics to improve the sensed signal quality (see Fig. \ref{fig:sensor_structure}). The converted understandable data   $S_{con}$ is typically an analog signal, which is given as an input to a differential amplifier to remove the common mode noise. A single or multiple stages of signal conditioning filters are applied afterward, which is followed by an analog-to-digital converter (ADC) for the data digitization. Filters are typically LPF used to remove high-frequency noise signals from the measurement. Typically, analog sensors output the analog signals from the differential amplifier or filters directly, while digital sensors contain the LPF and ADC. %Moreover, sensors can have a calibration block to calibrate the measurement before the digitization of the data.

\subsection{Operating region and saturation region}
\label{subsec:Operating region and saturation region}

\begin{figure}[ht!]
\centering
\includegraphics[width=0.18\textwidth,height=0.1\textheight]{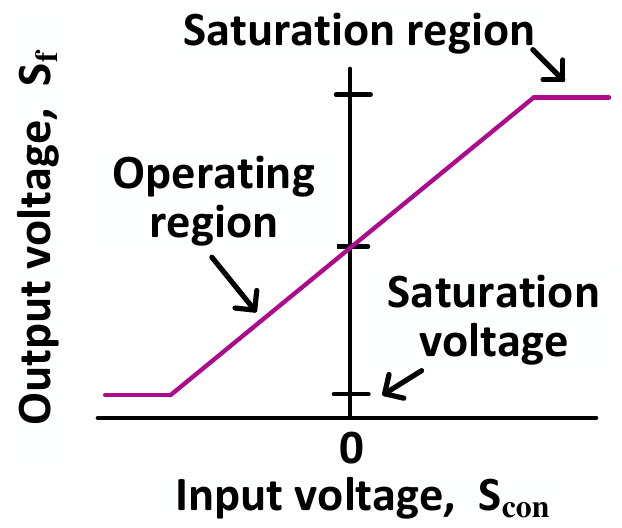}
\vspace{-1.30em}
\caption{Operating and saturation region of a sensor.}
\label{fig:saturation_region}
\vspace{-01.310em}
\end{figure}

As mentioned in Sections \ref{subsec:Sensor physics} and \ref{subsec:Sensor electronics}, the shape of the data $S_{con}$ from the transducer can be linear, exponential or any other types and is given as an input to different filtration blocks (see Fig. \ref{fig:sensor_structure}).   As these filters are powered by a finite power supply, the maximum value of $S_{con}$ the filters can handle is limited by the power supply. In fact, the output $S_f$ from the filtration block will begin to flatten when the $S_{con}$ is too large and the power supply limits are reached. The region, where the output $S_f$ is  not flattened, is known as the operating region of the sensor (see Fig. \ref{fig:saturation_region}). This region is typically linear between the signal $S_{con}$ and $S_f$. In contrast, the region, where the output $S_f$ is flattened, is known as the saturation region of the sensor. Please note that the exact value of the input $S_{con}$ cannot be recovered while the filter output $S_f$ is in the saturation region. 

\vspace{-1.00em}

\begin{table*}[ht!]
    \centering
    \footnotesize
    \caption{Summary of the notable published work on sensor attack. (In-band = fake and legitimate signal share same frequency band; Out-band = fake and legitimate signal share different frequency band)}
    \vspace{-01.1800em}
    \begin{tabular}{p{0.2cm} | p{0.7cm}| p{0.6cm}|l|l|>{\columncolor[rgb]{1,0.92,0.96}}l|>{\columncolor[rgb]{1,0.92,0.96}}l|>{\columncolor[rgb]{1,0.92,0.96}}p{2.050cm}}
    \hline
        \cellcolor [gray]{0.85}\textbf{Sl.} & \cellcolor [gray]{0.85}\textbf{Paper} & \cellcolor [gray]{0.85}\textbf{Year} & \cellcolor [gray]{0.85} \textbf{Sensor type} & \cellcolor [gray]{0.85} \textbf{Sensor} & \cellcolor [gray]{0.85} \textbf{Injected signal} & \cellcolor [gray]{0.85} \textbf{Injected signal pattern} & \cellcolor [gray]{0.85} \textbf{Industry} \\ 
        \hline
        \hline
        1 & \cite{barua2020hallspoofing}  & 2020  & Magnetic & Hall  & Magnetic field  & In-band, out-band  & Power grid \\ 
        \hline
        2 &  \cite{shoukry2013non} & 2013 & Magnetic  & Hall   &  Magnetic field & In-band & Automotive \\
        \hline
        3 &  \cite{wang2017sonic} & 2017  & Inertial & Accelerometer \& Gyroscope & Ultrasound  & Out-band resonant frequency  & AR/VR \\ 
        \hline
        4 & \cite{son2015rocking}  & 2015 & Inertial  & Gyroscope & Sound wave & Out-band resonant frequency  & Drone \\ 
        \hline
        5 & \cite{trippel2017walnut} & 2017 & Inertial  & Accelerometer & Ultrasound & Out-band resonant frequency & Smart device \\ 
        \hline
        6 & \cite{tu2018injected} & 2018  & Inertial & Accelerometer \& Gyroscope & Sound wave & Out-band resonant frequency & Smart device \\ 
        \hline
        7 & \cite{bolton2018blue} & 2018 & Inertial & Shock sensor &  Sound wave & Out-band & Hard disk  \\ 
        \hline
        8 &  \cite{barua2022pressuresensor} & 2022 & Pressure & Pressure sensor & Sound wave & Out-band resonant frequency &  NPR\\
        \hline
        9 &  \cite{tu2021transduction} & 2021 & Pressure & Pressure sensor & EMI & Out-band &  Inflation pump\\
        \hline
        10 &  \cite{park2016ain} & 2016 & Optical & Optical sensor & Infrared & Out-band &  Infusion pump\\
        \hline
        11 &  \cite{davidson2016controlling} & 2016 & Optical & Camera \& Lidar & Light & In-band &  UAV\\
        \hline
        12 &  \cite{shin2017illusion} & 2017 & Optical & Lidar & Light & In-band &  Automotive\\
        \hline
        13 &  \cite{zhang2017dolphinattack} & 2017 & Acoustic & Microphone & Ultrasound & Out-band &  Smart device\\
        \hline
        14 &  \cite{yan2016can} & 2016 & Acoustic & Ultrasound & Ultrasound & In-band &  Automotive\\
        \hline
        15 &  \cite{kune2013ghost} & 2013 & Analog &  Defibrillator & EMI & In-band &  Medical device\\
        \hline
        
    \end{tabular}
    \vspace{-1.20em}
    \label{table:Sensor attack}
\end{table*}

\subsection{Sensor physics from security perspective}
\label{subsec:Sensor physics from security perspective}

\textbf{Entry point for an attacker:} Over the last three decades, transducers and sensing structures have been technically improved in terms of stability, accuracy, and sensitivity; however, to the best of our knowledge, designers still do not consider security as one of the fundamental requirements while designing transducers and sensing structures of sensors. As a result, it is noteworthy that \textit{all} the transducers and sensing structures are naive, and they cannot differentiate between legitimate input signals and malicious fake input signals. As a consequence, if an adversary injects  malicious fake input signals into the sensing structure, the injected fake input signals are converted into understandable data by transducers and then propagate up to the system controller. The system controller blindly trusts signals coming from sensors. Therefore, this can be an entry point for an attacker. An attacker can noninvasively inject fake input signals to the transducers or sensing structures and can fool the system controller, resulting in a catastrophic  failure, system shutdown, or disruption of the system's normal behavior.

\textbf{Systematization of the attack signal:} We need to emphasize again that the \textit{only} reason that facilitates this attack is that the sensor structure and the transducer cannot differentiate between the legitimate input signals and the malicious fake input signals. Therefore, this type of attack is denoted by the \textit{transduction attack} \cite{fu2018risks, yan2020sok}. Let's denote malicious fake input signals  by $S^f_{in}$. Therefore, the output $S_{con}$ from the transducer in Eqn. \ref{eqn:sensor_equation} can be written as:

\vspace{-0.700em}
\begin{equation}
S^f_{con} = k \times f\{S_{in} + S^f_{in}\}
\label{eqn:sensor_equation_fake}
\end{equation}
\vspace{-1.0em}

where $S^f_{con}$ is the output from the transducer after the injection of the malicious fake input signals $S^f_{in}$. The term $S^f_{con}$ can propagate from the  transducer to the upper level as the existing sensor electronics blindly believe what is coming from the transducer.

\textbf{Scope of our work:} We consider only those attacks that originate from the fake signal injection into the sensing structures and transducers. Let us give an example to clarify the scope of our work. A pressure sensor can measure room pressure. If an attacker changes the room pressure by switching on/off the heating, ventilation, and air conditioning (HVAC) unit of the room, the room pressure changes. As a result, the pressure sensor measures a corrupted room pressure. We are not considering this type of attack on the pressure sensor. Instead, we are considering that type of attack, where the attacker directly injects fake signals into the sensing structure (i.e., diaphragm) and transducer of the pressure sensor. %To prevent this  transduction attack, a redesigning in the transducer and sensing structure are required.

\section{Sensor attack model}
\label{sec:Attack model}

\begin{figure}[ht!]
\centering
\includegraphics[width=0.45\textwidth,height=0.11\textheight]{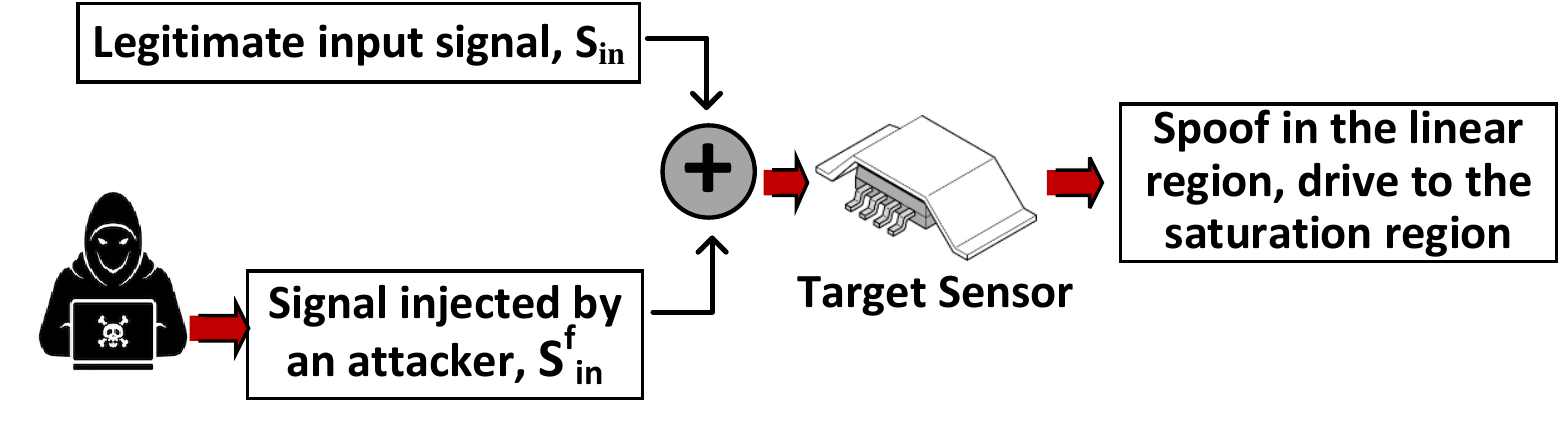}
\vspace{-1.30em}
\caption{Basic components of the sensor attack model.}
\label{fig:attack_model}
\vspace{-01.10em}
\end{figure}

The basic components of the sensor attack model is explained below and also shown in Fig. \ref{fig:attack_model}.

\textbf{i. Attacker's capability:} The attacker may not get a long time to modify the sensor like a lunch-time attack \cite{garman2016dancing}. Instead, the attacker may get a brief access near the sensor to inject the fake input signal $S^f_{in}$ from a \textit{close} distance. The fake input signal $S^f_{in}$ is a physical signal coming from the physical domain in different forms, such as acoustics, ultrasound, infrared, visible light, magnetic field, and electric field, and impacting the cyber domain \cite{chhetri2017cross} of the sensor and connected systems.

\textbf{ii. Noninvasive and stealthy attack:}  The attacker is not allowed to invasively access and modify any hardware and software of the sensor using physical tempering. This type of physical invasive attack is out of the scope of our study. Instead, our attack considers a scenario where the attacker injects a fake input signal to noninvasively perturb the sensor's transducer in a stealthy manner.

\vspace{-0.500em}

\textbf{iii. Attack's outcome:} As mentioned in Section \ref{subsec:Operating region and saturation region}, a sensor has a linear operating region and a flattened saturation region. The attacker can inject a fake input signal $S^f_{in}$ to spoof the sensor output in its linear region or drive the sensor output to its saturation region. 

In the linear region, the attacker can force the sensor to work at a particular operating point and can cause an adversarial control. For example, an attacker can use magnetic fields to spoof a Hall sensor in its linear region located in a solar inverter and can intentionally change the power of the solar inverter \cite{barua2020hallspoofing, Barua2019, barua2020special}.

In the saturation region, the input-output linear relationship of sensors gets flattened, and sensors go completely blind to any variation of the input. This may cause catastrophic failure in the normal operation of the connected systems resulting in a DoS attack. Typically, the attacker requires a stronger fake input signal to drive the sensor output to its saturation region \cite{barua2022premsat}. 

\begin{table*}[ht!]
    \centering
    \caption{Summary of the notable published work on sensor defense. (In-band = fake and legitimate signal share same frequency band; Out-band = fake and legitimate signal share different frequency band)}
    \vspace{-01.1800em}
    \footnotesize
    \begin{tabular}{p{0.2cm} | p{0.7cm}| p{0.6cm}|l|l|>{\columncolor[rgb]{
0.92,1,0.92}}l|>{\columncolor[rgb]{
0.92,1,0.92}}l|}
    \hline
        \cellcolor [gray]{0.85}\textbf{Sl.} & \cellcolor [gray]{0.85}\textbf{Paper} & \cellcolor [gray]{0.85}\textbf{Year} & \cellcolor [gray]{0.85} \textbf{Sensor type} & \cellcolor [gray]{0.85} \textbf{Sensor} & \cellcolor [gray]{0.85} \textbf{Defense technique} & \cellcolor [gray]{0.85} \textbf{Research challenge}  \\ 
        \hline
        \hline
        1 & \cite{barua2020hallspoofing}  & 2020  & Magnetic & Hall  & Shielding  & Not scalable, bulky  \\ 
        \hline
        2 &  \cite{shoukry2013non} & 2013 & Magnetic  & Hall   &  PyCRA & Can be bypassed  \\
        \hline
        3 &  \cite{wang2017sonic} & 2017  & Inertial & Accelerometer \& Gyroscope & Noise cancellation  & Not for in-band signal, not for saturation  \\ 
        \hline
        4 & \cite{son2015rocking}  & 2015 & Inertial  & Gyroscope & Sound wave & Not for in-band signal, not for saturation    \\ 
        \hline
        5 & \cite{trippel2017walnut} & 2017 & Inertial  & Accelerometer & Randomized and $180^0$ out-of-phase sampling & Not for in-band signal, DC attack signal  \\ 
        \hline
        6 & \cite{tu2018injected} & 2018  & Inertial & Accelerometer \& Gyroscope & LPF and dampening & Not for in-band signal, not for saturation \\ 
        \hline
        7 & \cite{bolton2018blue} & 2018 & Inertial & Shock sensor &  Sensor fusion & Costly, redundant, not for saturation   \\ 
        \hline
        8 &  \cite{barua2022pressuresensor} & 2022 & Pressure & Pressure sensor & LPF & Not for in-band signal \\
        \hline
        9 &  \cite{tu2021transduction} & 2021 & Pressure & Pressure sensor & Transduction shield & Finite physical distance, not for saturation \\
        \hline
        10 &  \cite{park2016ain} & 2016 & Optical & Optical sensor & PyCRA & Can be bypassed \\
        \hline
        11 &  \cite{davidson2016controlling} & 2016 & Optical & Camera \& Lidar & Modified optical flow algorithm & Not applicable \\
        \hline
        12 &  \cite{shin2017illusion} & 2017 & Optical & Lidar & Sensor fusion, redundancy & Costly, not for saturation \\
        \hline
        13 &  \cite{zhang2017dolphinattack} & 2017 & Acoustic & Microphone & Noise cancellation & Not for in-band signal \\
        \hline
        14 &  \cite{yan2016can} & 2016 & Acoustic & Ultrasound & Acoustic noise reduction (ANR) & For acoustic only, Not for out-band signal \\
        \hline
        15 &  \cite{kune2013ghost} & 2013 & Analog &  Defibrillator & Adaptive filter, LPF & Not for saturation\\
        \hline
        
    \end{tabular}
    \vspace{-1.20em}
    \label{table:Sensor defense}
\end{table*}

\vspace{-0.40em}
\section{Current Progress}
\label{sec:Current Progress}

Sensor security is comparatively a new domain in the security community that has  ongoing research for the last two decades. However, over the last decades, the research on sensor security has been accelerated due to the advent of the smart sensing systems in autonomous vehicles (AV), internet-of-things (IoT), and smart automation systems. Our paper broadly classifies the ongoing research on sensor security into two categories: (i) current progress on sensor attack and (ii) current progress on sensor defense.

\vspace{-0.40em}
\subsection{Current progress on sensor attack}
\label{subsec:Current Progress on sesnor attack}

We discuss sensor attacks in the following broad categories. A summary of the most notable published work on sensor attack is given in Table \ref{table:Sensor attack} with their publication years.

\textbf{Attack on magnetic sensors:} Barua et al. \cite{barua2020hallspoofing} demonstrated a noninvasive attack on Hall sensors located in a solar inverter using a magnetic field from a close distance, resulting in a shutdown of a weak micro-grid. Shoukry et al. \cite{shoukry2013non} showed a disruptive magnetic spoofing attack on a Hall sensor located in an anti-lock braking system (ABS) of a vehicle, resulting in a possible brake failure.

\textbf{Attack on inertial sensors:} Wang et al. \cite{wang2017sonic} used an ultrasonic gun to spoof different inertial sensors, such as MEMS accelerometers and gyroscopes, at their resonant frequencies to create havoc in the connected systems. Son et al. \cite{son2015rocking} used a powerful sound wave to spoof the gyroscope of a drone at its resonance frequency, making the drone uncontrollable. Trippel et al. \cite{trippel2017walnut}, and Tu et al. \cite{tu2018injected} showed an adversarial control over MEMS accelerometers and gyroscopes using acoustic signals at their resonant frequencies. Bolton et al. \cite{bolton2018blue} showed how acoustic signal can compromise the MEMS shock sensor located in a hard disk drive.

\textbf{Attack on pressure sensors:} Barua et al. \cite{barua2022pressuresensor} demonstrated a spoofing attack on pressure sensors located in a negative pressure room (NPR), using acoustic signal at the sensor's resonant frequency. Tu et al. \cite{tu2021transduction} showed a deliberate  electromagnetic interference (EMI) attack on an inflation pump's pressure sensor while impacting  the system's actuation.

\textbf{Attack on optical sensors:} Park et al. \cite{park2016ain} used infrared to spoof optical sensors of an infusion pump to deliver overdose to patients. Davidson et al. \cite{davidson2016controlling} reported how spoofing optical sensors of an unmanned aerial vehicle (UAV) can compromise its complete control. Shin et al. \cite{shin2017illusion} showed a spoofing attack on Lidar to create illusions of objects appearing closer in automotive systems.

\textbf{Attack on acoustic sensors:} Zhang et al. \cite{zhang2017dolphinattack} injected inaudible commands into a microphone using ultrasonic carriers. Yan et al. \cite{yan2016can} showed an attack on ultrasonic sensors of a vehicle using acoustic waves to impair vehicle safety.

\textbf{Attack on other analog sensors:} Kune et al. \cite{kune2013ghost} spoofed sensors by EMI to induce defibrillation shocks on cardiac devices.

It is just a matter of time before more attacks on sensors will emerge from different attack surfaces as sensors are getting complex and sophisticated nowadays without improving their security.

\vspace{-0.5em}
\subsection{Current progress on sensor defense}
\label{subsec:Current Progress on sesnor defense}

We discuss sensor defense in the following broad categories. A summary of the most notable sensor defenses is given in Table \ref{table:Sensor defense}.

\textbf{Defense for magnetic sensors:} Barua et al. \cite{barua2020hallspoofing} proposed strong magnetic shielding with a secure surrounding to prevent the magnetic spoofing attack on Hall sensors. Shoukry et al. \cite{shoukry2013non} proposed PyCRA to randomize the transmission and reception of signals to prevent magnetic spoofing attacks on ABS sensors of automotive. However, PyCRA only works for active sensors \cite{shoukry2013non}.

\textbf{Defense for inertial sensors:} Trippel et al. \cite{trippel2017walnut} proposed randomized and $180^0$ out-of-phase sampling to nullify acoustic spoofing signals injected into MEMS accelerometers. Son et al. \cite{son2015rocking} proposed resonance tuning using a feedback capacitor to prevent the out-band resonant frequency. Wang et al. \cite{wang2017sonic} proposed to use an external microphone to detect the resonating sound and perform noise cancellation. Tu et al. \cite{tu2018injected} proposed low-pass filtering (LPF) along with the dampening of the injected ultrasound.

\textbf{Defense for pressure sensors:} Barua et al. \cite{barua2022pressuresensor} used an LPF to filter out the sound wave from pressure sensors to prevent acoustic spoofing on pressure sensors. Tu et al. \cite{tu2021transduction} used a transduction shield to measure the fake signal first and then subtract it from the  corrupted signal to recover the legitimate signal. 

\textbf{Defense for optical sensors:} Park et al. \cite{park2016ain} measured the light intensity to detect the attack and used PyCRA to prevent it. Shin et al. \cite{shin2017illusion} propose sensor fusion and redundancy to prevent the optical attack on a lidar.

\textbf{Defense for acoustic sensors:} Zhang et al. \cite{zhang2017dolphinattack} used a modulation - demodulation based noise canceling technique to cancel out the injected ultrasound. Yan et al. \cite{yan2016can} proposed shielding and acoustic noise reduction (ANR) by emitting a sound with minor phase and amplitude adjustment.

\textbf{Defense for other analog sensors:} Kune et al. \cite{kune2013ghost} proposed adaptive filtering to estimate the spoofing attack signal first and then subtract the estimated attack signal from the original signal to clean up the original signal.

\vspace{-0.4em}
\section{Research challenges}
\label{sec:Research challenges}

The defense techniques described in Section \ref{subsec:Current Progress on sesnor defense} have research challenges. A summary of research challenges is given in Table \ref{table:Sensor defense}. 

\textbf{Shielding and dampening:} Barua et al. \cite{barua2020hallspoofing} and Tu et al. \cite{tu2018injected} proposed shielding to dampen the injected fake magnetic field and fake ultrasound, respectively. Though shielding is a cheap and quick countermeasure for a few of the fake injected signals, such as sound wave, ultrasound, and infrared, shielding might fail for other signals, such as magnetic fields. For example, a strong shield again may fail to a stronger magnetic field injected by an attacker. To increase the shielding property of a shield so that it can work for a stronger attack signal, the designer may need to increase the thickness of the shield. As even a thick shield can be penetrated by a stronger magnetic field, there is no \textit{sweet spot} for shielding to claim that it can prevent a magnetic field of any strength. Therefore, only shielding cannot be considered as a \textit{sole} defense for sensors.

\textbf{Randomization of transmission (PyCRA):}  Shoukry et al \cite{shoukry2015pycra} proposed PyCRA; however, PyCRA only works for active sensors, not for passive sensors. Moreover, Shin et al. \cite{shin2016sampling} showed that the  implemented  authentication mechanism of PyCRA can be successfully bypassed  with a low-cost circuit. This proves that there is currently no effective, robust and generalizable defense scheme against active sensor spoofing attacks.

\textbf{Randomized and $180^0$ out-of-phase sampling:} These two techniques from \cite{trippel2017walnut} only work for out-band resonant frequency attack signals. However, they do not work other than a specific resonant frequency, for example, any attack frequency. Moreover, they do not work against a DC forged signal because randomized sampling cannot filter out a DC signal. In addition, they do not work when the sensor output is flattened in the saturation region. 

\textbf{Adaptive filter and transduction shield:} Both of these defense techniques from \cite{kune2013ghost,tu2021transduction} work similarly by estimating the spoofing attack signal first and then subtracting the estimated attack signal from the original signal to clean up the original signal. These techniques will fail in the following two scenarios: (i) Because of the finite physical distance between the adaptive filter or transduction shield and the compromised sensor, the adaptive filter or transduction shield cannot measure the exact amplitude of the external attack signals. This is why we can not simply subtract the estimated attack signals from the original signals to recover the original signal. (ii) They do not work when the sensor output is flattened in the saturation region.

\textbf{LPF and other filtering techniques:} LPF and other filtering techniques in \cite{barua2022pressuresensor,kune2013ghost} only work as a successful defense when the fake attack signal has a predicable frequency spectrum and is an out-band signal. If the injected fake signal shares the same frequency band as the legitimate signal, LPF or other filtering techniques cannot accurately filter out the fake attack signals. The reason behind this is that while filtering the attack signals, they also filter out the same band of legitimate signals.

\textbf{Machine learning (ML) techniques:}  Machine learning (ML)  techniques  require complex computations to converge for  attack detection and recovery, requiring powerful hardware resources. Therefore, they are not suitable for low-power real-time sensor systems with constrained resources. In addition, they may not work against a time-varying magnetic spoofing as a time-varying signal may create oscillations between two safe states of the controller, and they are incapable of handling these oscillations in real-time.

\vspace{-0.54em}
\subsection{Area of focus}
\label{subsec:Bottleneck or summary of challenges}

%\textbf{Bottleneck:} 

If we analyze all the notable defenses from Table \ref{table:Sensor defense} published throughout the last decade, we can conclude the following points:

$\blacksquare$  The defenses were proposed on an ad-hoc basis. For example, the researchers first find a security issue with the sensor and then try to come up with a defense to overcome it. Researchers only focus on a specific security issue rather than focusing on all the corner cases from all the attack surfaces. Therefore, a proposed defense for a particular type of sensor is not applicable to other types. For example, the randomized and $180^0$ out-of-phase sampling \cite{trippel2017walnut}, which are applicable for MEMS inertial sensors to nullify out-band resonant attack frequency, is not applicable for magnetic Hall sensors because Hall sensors can have in-band attack frequency other than a specific out-band resonant frequency \cite{barua2020hallspoofing}.

$\blacksquare$  There is no defense technique exists in the literature that can prevent sensors from going into the saturation region or recover information from the sensor's saturation region.

$\blacksquare$  There is no defense, which can contain a fake injected signal having the same frequency as the legitimate signal.

$\blacksquare$  As PyCRA \cite{shoukry2015pycra} can be bypassed, there is no defense in the literature for active sensors till to date. 

Therefore, sensor security is still a \textit{malnourished} domain, which requires more attention from security researchers. 

\section{Future roadmap}
\label{sec:Future roadmap}

\textbf{Sensor heterogeneity:} The fundamental challenge of designing and modeling a robust and secured sensor is the heterogeneity of sensor types.  The heterogeneity of sensors exists because of the sensor's signal modality, which differs from the sensor to sensor. For example, a Hall sensor handles magnetic fields and a pressure sensor handles pressure waves as the signal modality.  As sensors differ from each other in terms of signal modality, the proper selection of transducers and sensor electronics also differ from sensor to sensor depending on the signal modality. Therefore, a single generalized defense technique that is applicable to all sensors will be quite complicated. However,  it is still worth giving it a try from a sensor security research point of view because the future days will be the days for smart sensors, and security will be a big concern for them.

\subsection{Golden reference for sensor defense}
\label{subsec:Golden reference for sensor defense}

%\textbf{Golden reference for sensor defense:} 
The next generation of research on sensor security should have to consider the security from designing the transducer to the implementation of the sensor electronics. Instead of implementing an ad-hoc defense, researchers must ensure a checklist before finally adopting a technique as a sensor defense. We name this checklist by the \textit{Golden reference} for a sensor defense against a transduction attack. This golden reference has the following points.

\begin{enumerate}[label=(\roman*),leftmargin=\parindent,labelwidth=\parindent,labelsep=4pt]
\vspace{0.0em}
\item  The defense should simultaneously work for in-band and out-band fake injected input signals.

\vspace{0.0em}
\item The defense should prevent a sensor  from going into the saturation region because of the injection of fake input signals.

\vspace{0.0em}
\item  The defense should contain a fake injected signal even it has the same frequency as the legitimate input signal.  

\vspace{0.0em}
\item  The defense should work for active and passive sensors. If  a single defense does not simultaneously work for both sensors, there should be a defense targeting an active sensor and another separate defense targeting a passive sensor.

\vspace{0.0em}
\item  The defense should be hard real-time.

\vspace{0.0em}
\item  The defense should not hamper the existing data processing speed or bandwidth of the sensor. 
 
\end{enumerate}

\subsection{Roadmap to achieve the Golden reference}
\label{subsec:Roadmap to achieve the Golden reference}

The Golden reference for sensor defense can not be achieved \textit{alone} by only hardware or only software modification; instead, a hardware/software (HW/SW) co-design approach is required in the sensor domain.  The sensors should be redesigned from the transducer level to the sensor electronics level. A smart transducer should be built instead of a naive one, and an intelligent and low-complexity algorithm should be adopted in the sensor electronics. We explain the roadmap to achieve the Golden reference below.

\subsubsection{\textbf{Encrypted analog signal:}}

%$\blacksquare$ \textbf{Encrypted analog signal and transducer:} 

The main reason for sensor vulnerability is that the legitimate analog signal $S_{in}$, which is going to be measured by the sensor,  is not encrypted before going into the transducer. Therefore, the attacker can use a fake signal $S^f_{in}$  to corrupt the legitimate signal $S_{in}$. This problem can be solved by encrypting the legitimate analog signal with a key in the \textit{analog domain} and decrypting the legitimate signal in the transducer side or in the sensor electronics using the same key. The analog domain encryption can be achieved using the following ways:

\textbf{$\blacksquare$ First}, an orthogonal noise can be padded with the legitimate signal $S_{in}$ to hide the information from the attack surface. This method is known as analog scrambling \cite{pichler1982analog}.

\textbf{$\blacksquare$ Second}, the legitimate signal $S_{in}$ can be mapped into the broad spectrum. This technique is known as frequency hopping spread spectrum \cite{ipatov2005spread} and can be adopted in the sensor domain.  

\textbf{$\blacksquare$ Third}, the legitimate signal $S_{in}$ can be encrypted in the analog domain using a pseudo-noise, which is  the original key. More keys can be generated using the original key to decrypt the signal \cite{slimani2018encryption}.

Please note that encryption of the legitimate analog  signal is the \textit{ultimate} solution for a robust sensor. This defense technique would achieve the points i-iv of the Golden reference in Section \ref{subsec:Golden reference for sensor defense}; however, it may or may not hamper the data processing speed (point v) and real-time requirement  (point vi) of the sensor depending upon the encryption complexity.

\subsubsection{\textbf{Modifying the transducer:}} 
\begin{figure}[ht!]
\centering
\includegraphics[width=0.25\textwidth,height=0.1\textheight]{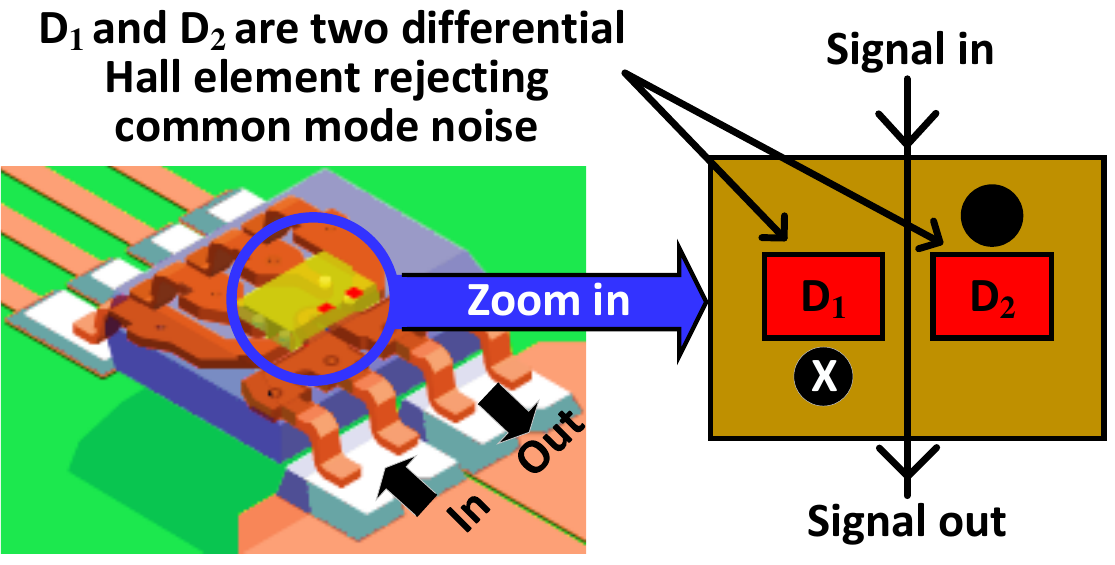}
\vspace{-1.3em}
\caption{A basic overview of the differential sensing.}
\label{fig:differential}
\vspace{-01.510em}
\end{figure}

The transducer must be resilient enough to reject the injected fake signal $S^f_{in}$. As a transducer is an entry point to the sensor, if a transducer can reject the fake signal, this approach would diminish the burden of using a complex encryption algorithm in the sensor hardware. Industry is adopting this scheme nowadays where ever it is feasible. For example, Hall current sensors use a Hall element as a transducer. Two Hall elements (see Fig. \ref{fig:differential}) are placed inside of a Hall current sensor  in a differential manner to reject common-mode noise \cite{diff_sensor_allegro}. As the injected fake signal $S^f_{in}$ is common to the differential Hall element, the  injected fake signal $S^f_{in}$ can be considered as common-mode noise and can be eliminated from the sensor at the transducer level. Though this strategy is quite novel, it has its own implementation challenge for all types of sensors. Research along this direction would be \textit{influential} in future for a secured sensor.   

\vspace{-1.00em}

\subsubsection{\textbf{Preventing sensor saturation:}} As mentioned in Section \ref{sec:Attack model}, if an attacker can drive the sensor to its saturation region, the output gets flattened, and no information can be retrieved, resulting in a DoS attack on sensors. This attack is known as saturation attack \cite{barua2022premsat}. The core idea behind preventing a saturation attack is to generate an internal signal, which has the \textit{same} strength but in \textit{opposite polarity} to the injected fake signal, so that the internal signal can nullify the injected fake signal. Barua et al. \cite{barua2022premsat} conceptualized and implemented this idea in the context of Hall magnetic sensors by providing a defense named \textit{PreMSat}. The idea of \textit{PreMSat} can be extended to other sensor types as well.

\vspace{-1.100em}
\begin{figure}[ht!]
\centering
\includegraphics[width=0.48\textwidth,height=0.091\textheight]{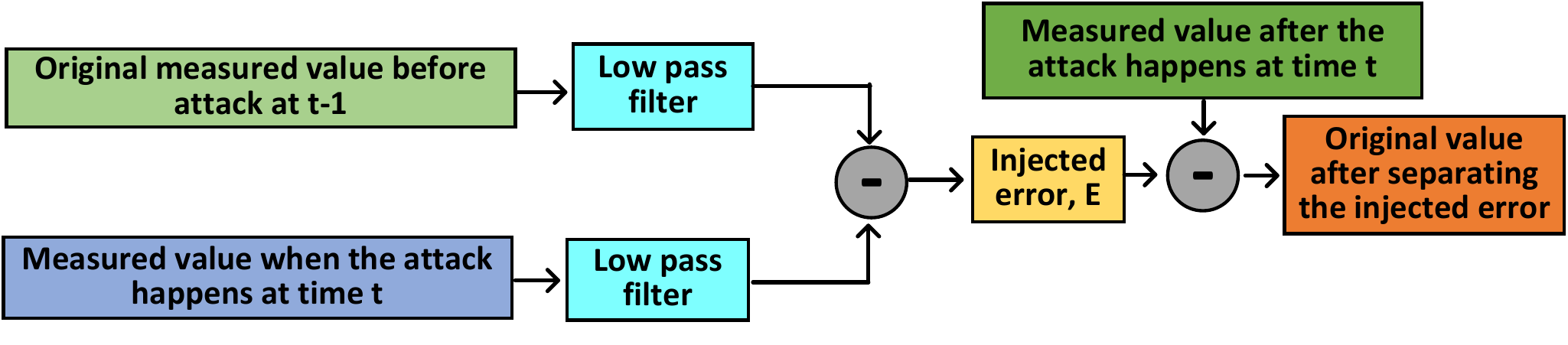}
\vspace{-2.5em}
\caption{A basic overview of the improved algorithm \cite{barua2022halc}.}
\label{fig:HALC}
\vspace{-01.310em}
\end{figure}

\subsubsection{\textbf{Improving the adaptive filter and transduction shield:}} The main bottleneck of using the adaptive filter and transduction shield is the introduced error in the estimated attack signals for the physical distance present between the sensor and the shield (see Section \ref{sec:Research challenges}). This problem can be solved by using an improved defense algorithm proposed by Barua et al. \cite{barua2022halc}. The main idea of this algorithm is that the difference between the measured signal during the attack and before the attack gives the amount of injected error after the attack. This algorithm tracks this difference all the time and generates a feedback signal to nullify the injected error by the attacker.  The idea is illustrated in Fig. \ref{fig:HALC}.

\subsubsection{\textbf{Context-aware anomaly detection:}} An  injected fake signal can be considered as an anomaly. Traditional anomaly detection algorithms are not low-power and hence, they are not suitable for sensor hardware. Therefore, a low-power and low- complexity algorithm may be another feasible option to run on the sensor hardware to detect injected fake signals. A low-power anomaly detection algorithm  named as Hierarchical Temporal Memory (HTM) can be evaluated to detect \textit{context aware} anomaly detection on the sensor data \cite{barua2020hierarchical, barua2020hierarchicalwip}. %HTM can be evaluated for anomaly detection and simultaneous data prediction during the transduction attack. HTM is an one-pass online learning algorithm and is faster compared to  traditional ML techniques. 
Moreover, the context-aware sensor association method \cite{yasaei2020iot} can be evaluated further as a defense technique for sensors. 

\subsubsection{\textbf{Control-theoretic approach:}} State estimation and state recovery based control-theoretic defense approaches can be another option to recover a system controller after the transduction attack.  Shoukry et al. \cite{shoukry2015secure} proposed reconstructing the sensor state to recover from a sensor spoofing attack using the satisfiability modulo theory (SMT). Wang et al. \cite{wang2014srid} demonstrated a graph-based technique to track states in the system controller to detect an intrusion. However, the correct direction will be to incorporate the  sensor's physics and physical knowledge of the system with the  control-theoretic approach for accurate estimation of the states. 

\subsubsection{\textbf{Sensor fusion based context aware:}} Sensor fusion and redundancy \cite{shin2017illusion, wang2017sonic, xu2018analyzing, bolton2018blue}  based approaches can be merged with ML algorithm to create  an appropriate context and abstraction of the sensor data. Therefore, during an attack, the sensor data can be recovered from a proper context using  abstracted sensor data. The data abstraction can be made intelligent and adaptive to tackle the continuous change of the sensor environment. However, sensor fusion adds extra price and complexity to the system; therefore, designers try to avoid this unless it is arguably required.

\section{Conclusion}
\label{sec:Conclusion}

In this paper, we present  a notion of sensor security, whereby, we focus on the importance of security measures that needs to be taken while designing a sensor. We discuss that sensors are vulnerable to external fake signals and give a summary of existing defenses in the sensor domain. We point out the limitations of existing defense techniques and emphasize that very little to no work exists in the sensor security domain. Therefore, we emphasize that the next generation of research on sensor security should have to ensure a Golden reference before finally adopting a technique as a sensor defense. We also provide a roadmap on how to achieve the Golden reference checklist while designing a robust and resilient sensor.

\begin{acks}

This work was partially supported by the National Science Foundation (NSF) under awards CMMI-1739503 and ECCS-2028269. Any opinions, findings, conclusions, or recommendations expressed in this paper are those of the authors and do not necessarily reflect the views of the funding agencies.

\end{acks}

%%
%% The next two lines define the bibliography style to be used, and
%% the bibliography file.
\Urlmuskip=0mu plus 1mu\relax
\bibliographystyle{ACM-Reference-Format}
\bibliography{bibfile}

%%% -*-BibTeX-*-
%%% Do NOT edit. File created by BibTeX with style
%%% ACM-Reference-Format-Journals [18-Jan-2012].

\begin{thebibliography}{35}

%%% ====================================================================
%%% NOTE TO THE USER: you can override these defaults by providing
%%% customized versions of any of these macros before the \bibliography
%%% command.  Each of them MUST provide its own final punctuation,
%%% except for \shownote{}, \showDOI{}, and \showURL{}.  The latter two
%%% do not use final punctuation, in order to avoid confusing it with
%%% the Web address.
%%%
%%% To suppress output of a particular field, define its macro to expand
%%% to an empty string, or better, \unskip, like this:
%%%
%%% \newcommand{\showDOI}[1]{\unskip}   % LaTeX syntax
%%%
%%% \def \showDOI #1{\unskip}           % plain TeX syntax
%%%
%%% ====================================================================

\ifx \showCODEN    \undefined \def \showCODEN     #1{\unskip}     \fi
\ifx \showDOI      \undefined \def \showDOI       #1{#1}\fi
\ifx \showISBNx    \undefined \def \showISBNx     #1{\unskip}     \fi
\ifx \showISBNxiii \undefined \def \showISBNxiii  #1{\unskip}     \fi
\ifx \showISSN     \undefined \def \showISSN      #1{\unskip}     \fi
\ifx \showLCCN     \undefined \def \showLCCN      #1{\unskip}     \fi
\ifx \shownote     \undefined \def \shownote      #1{#1}          \fi
\ifx \showarticletitle \undefined \def \showarticletitle #1{#1}   \fi
\ifx \showURL      \undefined \def \showURL       {\relax}        \fi
% The following commands are used for tagged output and should be
% invisible to TeX
\providecommand\bibfield[2]{#2}
\providecommand\bibinfo[2]{#2}
\providecommand\natexlab[1]{#1}
\providecommand\showeprint[2][]{arXiv:#2}

\bibitem[Alexander(2019)]%
        {diff_sensor_allegro}
\bibfield{author}{\bibinfo{person}{Latham Alexander}.}
  \bibinfo{year}{2019}\natexlab{}.
\newblock \bibinfo{booktitle}{\emph{{Common Mode Field Rejection in Coreless
  Hall-Effect Current Sensor ICs}}}.
\newblock
\newblock
\shownote{\url{https://www.allegromicro.com/-/media/files/application-notes/an269123-common-mode-field-rejection-in-coreless-current-sensor-ics.pdf}.
  (Accessed: 03-22-2022)}.


\bibitem[Barua and Al~Faruque(2019)]%
        {Barua2019}
\bibfield{author}{\bibinfo{person}{Anomadarshi Barua} {and}
  \bibinfo{person}{Mohammad~Abdullah Al~Faruque}.}
  \bibinfo{year}{2019}\natexlab{}.
\newblock \bibinfo{booktitle}{\emph{The Hall Sensor Security}}.
\newblock \bibinfo{publisher}{Springer Berlin Heidelberg},
  \bibinfo{address}{Berlin, Heidelberg}, \bibinfo{pages}{1--4}.
\newblock
\showISBNx{978-3-642-27739-9}
\urldef\tempurl%
\url{https://doi.org/10.1007/978-3-642-27739-9_1652-1}
\showDOI{\tempurl}


\bibitem[Barua and Al~Faruque(2020a)]%
        {barua2020hallspoofing}
\bibfield{author}{\bibinfo{person}{Anomadarshi Barua} {and}
  \bibinfo{person}{Mohammad~Abdullah Al~Faruque}.}
  \bibinfo{year}{2020}\natexlab{a}.
\newblock \showarticletitle{{Hall Spoofing: A Non-Invasive DoS Attack on
  Grid-Tied Solar Inverter}}. In \bibinfo{booktitle}{\emph{29th {USENIX}
  Security Symposium ({USENIX} Security 20)}}. \bibinfo{publisher}{{USENIX}
  Association}, \bibinfo{pages}{1273--1290}.
\newblock
\showISBNx{978-1-939133-17-5}
\urldef\tempurl%
\url{https://www.usenix.org/conference/usenixsecurity20/presentation/barua}
\showURL{%
\tempurl}


\bibitem[Barua and Al~Faruque(2020b)]%
        {barua2020special}
\bibfield{author}{\bibinfo{person}{Anomadarshi Barua} {and}
  \bibinfo{person}{Mohammad~Abdullah Al~Faruque}.}
  \bibinfo{year}{2020}\natexlab{b}.
\newblock \showarticletitle{Special session: Noninvasive sensor-spoofing
  attacks on embedded and cyber-physical systems}. In
  \bibinfo{booktitle}{\emph{2020 IEEE 38th International Conference on Computer
  Design (ICCD)}}. IEEE, \bibinfo{pages}{45--48}.
\newblock


\bibitem[Barua and Al~Faruque(2022a)]%
        {barua2022pressuresensor}
\bibfield{author}{\bibinfo{person}{Anomadarshi Barua} {and}
  \bibinfo{person}{Mohammad~Abdullah Al~Faruque}.}
  \bibinfo{year}{2022}\natexlab{a}.
\newblock \showarticletitle{{A Wolf in Sheep’s Clothing: Spreading Deadly
  Pathogens Under the Disguise of Popular Music}}. In
  \bibinfo{booktitle}{\emph{29th ACM Conference on Computer and Communications
  Security (CCS)}}.
\newblock


\bibitem[Barua and Al~Faruque(2022b)]%
        {barua2022halc}
\bibfield{author}{\bibinfo{person}{Anomadarshi Barua} {and}
  \bibinfo{person}{Mohammad~Abdullah Al~Faruque}.}
  \bibinfo{year}{2022}\natexlab{b}.
\newblock \showarticletitle{{HALC: A Real-time In-sensor Defense against the
  Magnetic Spoofing Attack on Hall Sensors}}. In \bibinfo{booktitle}{\emph{25th
  International Symposium on Research in Attacks, Intrusions and Defenses (RAID
  2022)}}.
\newblock


\bibitem[Barua and Al~Faruque(2022c)]%
        {barua2022premsat}
\bibfield{author}{\bibinfo{person}{Anomadarshi Barua} {and}
  \bibinfo{person}{Mohammad~Abdullah Al~Faruque}.}
  \bibinfo{year}{2022}\natexlab{c}.
\newblock \showarticletitle{{PreMSat: Preventing Magnetic Saturation Attack on
  Hall Sensors}}. In \bibinfo{booktitle}{\emph{International Conference on
  Cryptographic Hardware and Embedded Systems (TCHES 2022)}}.
\newblock


\bibitem[Barua et~al\mbox{.}(2020a)]%
        {barua2020hierarchicalwip}
\bibfield{author}{\bibinfo{person}{Anomadarshi Barua}, \bibinfo{person}{Deepan
  Muthirayan}, \bibinfo{person}{Pramod~P Khargonekar}, {and}
  \bibinfo{person}{Mohammad~Abdullah Al~Faruque}.}
  \bibinfo{year}{2020}\natexlab{a}.
\newblock \showarticletitle{Hierarchical temporal memory based machine learning
  for real-time, unsupervised anomaly detection in smart grid: WiP abstract}.
  In \bibinfo{booktitle}{\emph{2020 ACM/IEEE 11th International Conference on
  Cyber-Physical Systems (ICCPS)}}. IEEE, \bibinfo{pages}{188--189}.
\newblock


\bibitem[Barua et~al\mbox{.}(2020b)]%
        {barua2020hierarchical}
\bibfield{author}{\bibinfo{person}{Anomadarshi Barua}, \bibinfo{person}{Deepan
  Muthirayan}, \bibinfo{person}{Pramod~P Khargonekar}, {and}
  \bibinfo{person}{Mohammad~Abdullah Al~Faruque}.}
  \bibinfo{year}{2020}\natexlab{b}.
\newblock \showarticletitle{Hierarchical temporal memory based one-pass
  learning for real-time anomaly detection and simultaneous data prediction in
  smart grids}.
\newblock \bibinfo{journal}{\emph{IEEE Transactions on Dependable and Secure
  Computing}} (\bibinfo{year}{2020}).
\newblock


\bibitem[Bolton et~al\mbox{.}(2018)]%
        {bolton2018blue}
\bibfield{author}{\bibinfo{person}{Connor Bolton}, \bibinfo{person}{Sara
  Rampazzi}, \bibinfo{person}{Chaohao Li}, \bibinfo{person}{Andrew Kwong},
  \bibinfo{person}{Wenyuan Xu}, {and} \bibinfo{person}{Kevin Fu}.}
  \bibinfo{year}{2018}\natexlab{}.
\newblock \showarticletitle{Blue note: How intentional acoustic interference
  damages availability and integrity in hard disk drives and operating
  systems}. In \bibinfo{booktitle}{\emph{2018 IEEE Symposium on Security and
  Privacy (SP)}}. IEEE, \bibinfo{pages}{1048--1062}.
\newblock


\bibitem[Chhetri et~al\mbox{.}(2017)]%
        {chhetri2017cross}
\bibfield{author}{\bibinfo{person}{Sujit~Rokka Chhetri}, \bibinfo{person}{Jiang
  Wan}, {and} \bibinfo{person}{Mohammad~Abdullah Al~Faruque}.}
  \bibinfo{year}{2017}\natexlab{}.
\newblock \showarticletitle{Cross-domain security of cyber-physical systems}.
  In \bibinfo{booktitle}{\emph{2017 22nd Asia and South Pacific design
  automation conference (ASP-DAC)}}. IEEE, \bibinfo{pages}{200--205}.
\newblock


\bibitem[Davidson et~al\mbox{.}(2016)]%
        {davidson2016controlling}
\bibfield{author}{\bibinfo{person}{Drew Davidson}, \bibinfo{person}{Hao Wu},
  \bibinfo{person}{Rob Jellinek}, \bibinfo{person}{Vikas Singh}, {and}
  \bibinfo{person}{Thomas Ristenpart}.} \bibinfo{year}{2016}\natexlab{}.
\newblock \showarticletitle{{Controlling UAVs with sensor input spoofing
  attacks}}. In \bibinfo{booktitle}{\emph{10th $\{$USENIX$\}$ Workshop on
  Offensive Technologies ($\{$WOOT$\}$ 16)}}.
\newblock


\bibitem[Fu and Xu(2018)]%
        {fu2018risks}
\bibfield{author}{\bibinfo{person}{Kevin Fu} {and} \bibinfo{person}{Wenyuan
  Xu}.} \bibinfo{year}{2018}\natexlab{}.
\newblock \showarticletitle{Risks of trusting the physics of sensors}.
\newblock \bibinfo{journal}{\emph{Commun. ACM}} \bibinfo{volume}{61},
  \bibinfo{number}{2} (\bibinfo{year}{2018}), \bibinfo{pages}{20--23}.
\newblock


\bibitem[Garman et~al\mbox{.}(2016)]%
        {garman2016dancing}
\bibfield{author}{\bibinfo{person}{Christina Garman}, \bibinfo{person}{Matthew
  Green}, \bibinfo{person}{Gabriel Kaptchuk}, \bibinfo{person}{Ian Miers},
  {and} \bibinfo{person}{Michael Rushanan}.} \bibinfo{year}{2016}\natexlab{}.
\newblock \showarticletitle{Dancing on the lip of the volcano: Chosen
  ciphertext attacks on apple imessage}. In \bibinfo{booktitle}{\emph{25th
  $\{$USENIX$\}$ Security Symposium ($\{$USENIX$\}$ Security 16)}}.
  \bibinfo{pages}{655--672}.
\newblock


\bibitem[Ipatov(2005)]%
        {ipatov2005spread}
\bibfield{author}{\bibinfo{person}{Valeri~P Ipatov}.}
  \bibinfo{year}{2005}\natexlab{}.
\newblock \bibinfo{booktitle}{\emph{Spread spectrum and CDMA: principles and
  applications}}.
\newblock \bibinfo{publisher}{John Wiley \& Sons}.
\newblock


\bibitem[Kune et~al\mbox{.}(2013)]%
        {kune2013ghost}
\bibfield{author}{\bibinfo{person}{Denis~Foo Kune}, \bibinfo{person}{John
  Backes}, \bibinfo{person}{Shane~S Clark}, \bibinfo{person}{Daniel Kramer},
  \bibinfo{person}{Matthew Reynolds}, \bibinfo{person}{Kevin Fu},
  \bibinfo{person}{Yongdae Kim}, {and} \bibinfo{person}{Wenyuan Xu}.}
  \bibinfo{year}{2013}\natexlab{}.
\newblock \showarticletitle{Ghost talk: Mitigating EMI signal injection attacks
  against analog sensors}. In \bibinfo{booktitle}{\emph{2013 IEEE Symposium on
  Security and Privacy}}. IEEE, \bibinfo{pages}{145--159}.
\newblock


\bibitem[Park et~al\mbox{.}(2016)]%
        {park2016ain}
\bibfield{author}{\bibinfo{person}{Youngseok Park}, \bibinfo{person}{Yunmok
  Son}, \bibinfo{person}{Hocheol Shin}, \bibinfo{person}{Dohyun Kim}, {and}
  \bibinfo{person}{Yongdae Kim}.} \bibinfo{year}{2016}\natexlab{}.
\newblock \showarticletitle{{This ain't your dose: Sensor spoofing attack on
  medical infusion pump}}. In \bibinfo{booktitle}{\emph{10th $\{$USENIX$\}$
  Workshop on Offensive Technologies ($\{$WOOT$\}$ 16)}}.
\newblock


\bibitem[Pichler(1982)]%
        {pichler1982analog}
\bibfield{author}{\bibinfo{person}{Franz Pichler}.}
  \bibinfo{year}{1982}\natexlab{}.
\newblock \showarticletitle{Analog scrambling by the general fast Fourier
  transform}. In \bibinfo{booktitle}{\emph{Workshop on Cryptography}}.
  Springer, \bibinfo{pages}{173--178}.
\newblock


\bibitem[Shin et~al\mbox{.}(2017)]%
        {shin2017illusion}
\bibfield{author}{\bibinfo{person}{Hocheol Shin}, \bibinfo{person}{Dohyun Kim},
  \bibinfo{person}{Yujin Kwon}, {and} \bibinfo{person}{Yongdae Kim}.}
  \bibinfo{year}{2017}\natexlab{}.
\newblock \showarticletitle{Illusion and dazzle: Adversarial optical channel
  exploits against lidars for automotive applications}. In
  \bibinfo{booktitle}{\emph{International Conference on Cryptographic Hardware
  and Embedded Systems}}. Springer, \bibinfo{pages}{445--467}.
\newblock


\bibitem[Shin et~al\mbox{.}(2016)]%
        {shin2016sampling}
\bibfield{author}{\bibinfo{person}{Hocheol Shin}, \bibinfo{person}{Yunmok Son},
  \bibinfo{person}{Youngseok Park}, \bibinfo{person}{Yujin Kwon}, {and}
  \bibinfo{person}{Yongdae Kim}.} \bibinfo{year}{2016}\natexlab{}.
\newblock \showarticletitle{Sampling Race: Bypassing $\{$Timing-Based$\}$
  Analog Active Sensor Spoofing Detection on $\{$Analog-Digital$\}$ Systems}.
  In \bibinfo{booktitle}{\emph{10th USENIX Workshop on Offensive Technologies
  (WOOT 16)}}.
\newblock


\bibitem[Shoukry et~al\mbox{.}(2013)]%
        {shoukry2013non}
\bibfield{author}{\bibinfo{person}{Yasser Shoukry}, \bibinfo{person}{Paul
  Martin}, \bibinfo{person}{Paulo Tabuada}, {and} \bibinfo{person}{Mani
  Srivastava}.} \bibinfo{year}{2013}\natexlab{}.
\newblock \showarticletitle{Non-invasive spoofing attacks for anti-lock braking
  systems}. In \bibinfo{booktitle}{\emph{International Conference on
  Cryptographic Hardware and Embedded Systems}}. Springer,
  \bibinfo{pages}{55--72}.
\newblock


\bibitem[Shoukry et~al\mbox{.}(2015a)]%
        {shoukry2015pycra}
\bibfield{author}{\bibinfo{person}{Yasser Shoukry}, \bibinfo{person}{Paul
  Martin}, \bibinfo{person}{Yair Yona}, \bibinfo{person}{Suhas Diggavi}, {and}
  \bibinfo{person}{Mani Srivastava}.} \bibinfo{year}{2015}\natexlab{a}.
\newblock \showarticletitle{Pycra: Physical challenge-response authentication
  for active sensors under spoofing attacks}. In
  \bibinfo{booktitle}{\emph{Proceedings of the 22nd ACM SIGSAC Conference on
  Computer and Communications Security}}. \bibinfo{pages}{1004--1015}.
\newblock


\bibitem[Shoukry et~al\mbox{.}(2015b)]%
        {shoukry2015secure}
\bibfield{author}{\bibinfo{person}{Yasser Shoukry}, \bibinfo{person}{Pierluigi
  Nuzzo}, \bibinfo{person}{Nicola Bezzo}, \bibinfo{person}{Alberto~L
  Sangiovanni-Vincentelli}, \bibinfo{person}{Sanjit~A Seshia}, {and}
  \bibinfo{person}{Paulo Tabuada}.} \bibinfo{year}{2015}\natexlab{b}.
\newblock \showarticletitle{Secure state reconstruction in differentially flat
  systems under sensor attacks using satisfiability modulo theory solving}. In
  \bibinfo{booktitle}{\emph{2015 54th IEEE Conference on Decision and Control
  (CDC)}}. IEEE, \bibinfo{pages}{3804--3809}.
\newblock


\bibitem[Slimani and Merazka(2018)]%
        {slimani2018encryption}
\bibfield{author}{\bibinfo{person}{Dalila Slimani} {and}
  \bibinfo{person}{Fatiha Merazka}.} \bibinfo{year}{2018}\natexlab{}.
\newblock \showarticletitle{Encryption of speech signal with multiple secret
  keys}.
\newblock \bibinfo{journal}{\emph{Procedia computer science}}
  \bibinfo{volume}{128} (\bibinfo{year}{2018}), \bibinfo{pages}{79--88}.
\newblock


\bibitem[Son et~al\mbox{.}(2015)]%
        {son2015rocking}
\bibfield{author}{\bibinfo{person}{Yunmok Son}, \bibinfo{person}{Hocheol Shin},
  \bibinfo{person}{Dongkwan Kim}, \bibinfo{person}{Youngseok Park},
  \bibinfo{person}{Juhwan Noh}, \bibinfo{person}{Kibum Choi},
  \bibinfo{person}{Jungwoo Choi}, {and} \bibinfo{person}{Yongdae Kim}.}
  \bibinfo{year}{2015}\natexlab{}.
\newblock \showarticletitle{Rocking drones with intentional sound noise on
  gyroscopic sensors}. In \bibinfo{booktitle}{\emph{24th USENIX Security
  Symposium (USENIX Security 15)}}. \bibinfo{pages}{881--896}.
\newblock


\bibitem[Trippel et~al\mbox{.}(2017)]%
        {trippel2017walnut}
\bibfield{author}{\bibinfo{person}{Timothy Trippel}, \bibinfo{person}{Ofir
  Weisse}, \bibinfo{person}{Wenyuan Xu}, \bibinfo{person}{Peter Honeyman},
  {and} \bibinfo{person}{Kevin Fu}.} \bibinfo{year}{2017}\natexlab{}.
\newblock \showarticletitle{WALNUT: Waging doubt on the integrity of MEMS
  accelerometers with acoustic injection attacks}. In
  \bibinfo{booktitle}{\emph{2017 IEEE European symposium on security and
  privacy (EuroS\&P)}}. IEEE, \bibinfo{pages}{3--18}.
\newblock


\bibitem[Tu et~al\mbox{.}(2018)]%
        {tu2018injected}
\bibfield{author}{\bibinfo{person}{Yazhou Tu}, \bibinfo{person}{Zhiqiang Lin},
  \bibinfo{person}{Insup Lee}, {and} \bibinfo{person}{Xiali Hei}.}
  \bibinfo{year}{2018}\natexlab{}.
\newblock \showarticletitle{Injected and delivered: Fabricating implicit
  control over actuation systems by spoofing inertial sensors}. In
  \bibinfo{booktitle}{\emph{27th USENIX Security Symposium (USENIX Security
  18)}}. \bibinfo{pages}{1545--1562}.
\newblock


\bibitem[Tu et~al\mbox{.}(2021)]%
        {tu2021transduction}
\bibfield{author}{\bibinfo{person}{Yazhou Tu}, \bibinfo{person}{Vijay~Srinivas
  Tida}, \bibinfo{person}{Zhongqi Pan}, {and} \bibinfo{person}{Xiali Hei}.}
  \bibinfo{year}{2021}\natexlab{}.
\newblock \showarticletitle{Transduction Shield: A Low-Complexity Method to
  Detect and Correct the Effects of EMI Injection Attacks on Sensors}. In
  \bibinfo{booktitle}{\emph{Proceedings of the 2021 ACM Asia Conference on
  Computer and Communications Security}}. \bibinfo{pages}{901--915}.
\newblock


\bibitem[Wang et~al\mbox{.}(2014)]%
        {wang2014srid}
\bibfield{author}{\bibinfo{person}{Yong Wang}, \bibinfo{person}{Zhaoyan Xu},
  \bibinfo{person}{Jialong Zhang}, \bibinfo{person}{Lei Xu},
  \bibinfo{person}{Haopei Wang}, {and} \bibinfo{person}{Guofei Gu}.}
  \bibinfo{year}{2014}\natexlab{}.
\newblock \showarticletitle{Srid: State relation based intrusion detection for
  false data injection attacks in scada}. In \bibinfo{booktitle}{\emph{European
  Symposium on Research in Computer Security}}. Springer,
  \bibinfo{pages}{401--418}.
\newblock


\bibitem[Wang et~al\mbox{.}(2017)]%
        {wang2017sonic}
\bibfield{author}{\bibinfo{person}{Zhengbo Wang}, \bibinfo{person}{Kang Wang},
  \bibinfo{person}{Bo Yang}, \bibinfo{person}{Shangyuan Li}, {and}
  \bibinfo{person}{Aimin Pan}.} \bibinfo{year}{2017}\natexlab{}.
\newblock \showarticletitle{Sonic gun to smart devices: Your devices lose
  control under ultrasound/sound}.
\newblock \bibinfo{journal}{\emph{Black Hat USA}} (\bibinfo{year}{2017}),
  \bibinfo{pages}{1--50}.
\newblock


\bibitem[Xu et~al\mbox{.}(2018)]%
        {xu2018analyzing}
\bibfield{author}{\bibinfo{person}{Wenyuan Xu}, \bibinfo{person}{Chen Yan},
  \bibinfo{person}{Weibin Jia}, \bibinfo{person}{Xiaoyu Ji}, {and}
  \bibinfo{person}{Jianhao Liu}.} \bibinfo{year}{2018}\natexlab{}.
\newblock \showarticletitle{Analyzing and enhancing the security of ultrasonic
  sensors for autonomous vehicles}.
\newblock \bibinfo{journal}{\emph{IEEE Internet of Things Journal}}
  \bibinfo{volume}{5}, \bibinfo{number}{6} (\bibinfo{year}{2018}),
  \bibinfo{pages}{5015--5029}.
\newblock


\bibitem[Yan et~al\mbox{.}(2020)]%
        {yan2020sok}
\bibfield{author}{\bibinfo{person}{Chen Yan}, \bibinfo{person}{Hocheol Shin},
  \bibinfo{person}{Connor Bolton}, \bibinfo{person}{Wenyuan Xu},
  \bibinfo{person}{Yongdae Kim}, {and} \bibinfo{person}{Kevin Fu}.}
  \bibinfo{year}{2020}\natexlab{}.
\newblock \showarticletitle{Sok: A minimalist approach to formalizing analog
  sensor security}. In \bibinfo{booktitle}{\emph{2020 IEEE Symposium on
  Security and Privacy (SP)}}. IEEE, \bibinfo{pages}{233--248}.
\newblock


\bibitem[Yan et~al\mbox{.}(2016)]%
        {yan2016can}
\bibfield{author}{\bibinfo{person}{Chen Yan}, \bibinfo{person}{Wenyuan Xu},
  {and} \bibinfo{person}{Jianhao Liu}.} \bibinfo{year}{2016}\natexlab{}.
\newblock \showarticletitle{{Can you trust autonomous vehicles: Contactless
  attacks against sensors of self-driving vehicle}}.
\newblock \bibinfo{journal}{\emph{DEF CON}} \bibinfo{volume}{24},
  \bibinfo{number}{8} (\bibinfo{year}{2016}), \bibinfo{pages}{109}.
\newblock


\bibitem[Yasaei et~al\mbox{.}(2020)]%
        {yasaei2020iot}
\bibfield{author}{\bibinfo{person}{Rozhin Yasaei}, \bibinfo{person}{Felix
  Hernandez}, {and} \bibinfo{person}{Mohammad~Abdullah Al~Faruque}.}
  \bibinfo{year}{2020}\natexlab{}.
\newblock \showarticletitle{Iot-cad: context-aware adaptive anomaly detection
  in iot systems through sensor association}. In \bibinfo{booktitle}{\emph{2020
  IEEE/ACM International Conference On Computer Aided Design (ICCAD)}}. IEEE,
  \bibinfo{pages}{1--9}.
\newblock


\bibitem[Zhang et~al\mbox{.}(2017)]%
        {zhang2017dolphinattack}
\bibfield{author}{\bibinfo{person}{Guoming Zhang}, \bibinfo{person}{Chen Yan},
  \bibinfo{person}{Xiaoyu Ji}, \bibinfo{person}{Tianchen Zhang},
  \bibinfo{person}{Taimin Zhang}, {and} \bibinfo{person}{Wenyuan Xu}.}
  \bibinfo{year}{2017}\natexlab{}.
\newblock \showarticletitle{Dolphinattack: Inaudible voice commands}. In
  \bibinfo{booktitle}{\emph{Proceedings of the 2017 ACM SIGSAC Conference on
  Computer and Communications Security}}. \bibinfo{pages}{103--117}.
\newblock


\end{thebibliography}

\end{document}